\documentstyle[prd,aps,preprint,tighten,epsfig]{revtex}

\begin{document}

\draft

\title{Leptonic Unitarity Triangles in Matter}
\author{{\bf He Zhang} ~ and ~ {\bf Zhi-zhong Xing}}
\address{CCAST (World Laboratory), P.O. Box 8730, Beijing 100080, China \\
and Institute of High Energy Physics, Chinese Academy of Sciences, \\
P.O. Box 918 (4), Beijing 100049, China \\
({\it Electronic address: zhanghe@mail.ihep.ac.cn,
xingzz@mail.ihep.ac.cn}) }
\maketitle

\begin{abstract}
We present a geometric description of lepton flavor mixing and CP
violation in matter by using the language of leptonic unitarity
triangles. The exact analytical relations for both sides and inner
angles are established between every unitarity triangle in vacuum and
its effective counterpart in matter. The typical shape evolution of six
triangles with the terrestrial matter density is illustrated for a
realistic long-baseline neutrino oscillation experiment.
\end{abstract}
\newpage

\section{Introduction}

Recent solar \cite{SNO}, atmospheric \cite{SK}, reactor
(KamLAND \cite{KM} and CHOOZ \cite{CHOOZ}) and accelerator
(K2K \cite{K2K}) neutrino oscillation experiments have provided
us with very convincing evidence that neutrinos are massive
and lepton flavors are mixed. In the framework of three lepton families,
the phenomena of flavor mixing and CP violation are described by a
$3\times 3$ unitary matrix $V$, which relates the mass eigenstates
of three neutrinos $(\nu_1, \nu_2, \nu_3)$ to their flavor eigenstates
$(\nu_e, \nu_\mu, \nu_\tau)$:
\begin{equation}
\left ( \matrix{
\nu_e \cr
\nu_\mu \cr
\nu_\tau \cr} \right )
= \left ( \matrix{
V_{e1}  & V_{e2} & V_{e3} \cr
V_{\mu 1} & V_{\mu 2} & V_{\mu 3} \cr
V_{\tau 1} & V_{\tau 2} & V_{\tau 3} \cr} \right )
\left ( \matrix{
\nu_1 \cr
\nu_2 \cr
\nu_3 \cr} \right ) \; .
\end{equation}
No matter whether neutrinos are Dirac or Majorana particles,
leptonic CP and T violation in normal neutrino-neutrino or
antineutrino-antineutrino oscillations depends only upon a single
rephasing-invariant parameter ${\cal J}$ \cite{Jarlskog},
defined through
\begin{equation}
{\rm Im} ( V_{\alpha i} V_{\beta j} V^*_{\alpha j}
V^*_{\beta i} ) = {\cal J} \sum_{\gamma, k}
( \epsilon_{\alpha\beta\gamma} \epsilon_{ijk} ) \; ,
\end{equation}
where Greek and Latin subscripts run respectively
over $(e, \mu, \tau)$ and $(1,2,3)$. A challenging task of the future
long-baseline neutrino oscillation experiments is to measure ${\cal J}$,
so as to establish the existence of CP violation in the lepton sector.
In principle, a determination of ${\cal J}$ is also possible from four
independent moduli $|V_{\alpha i}|$ \cite{Xreview}, whose magnitudes
can be measured in some CP-conserving processes.

The unitarity of $V$ implies that its nine matrix elements are constrained
by two sets of normalization conditions and two sets of orthogonality
relations:
\begin{eqnarray}
\sum_i \left (V^*_{\alpha i} V_{\beta i} \right ) & = & ~
\delta_{\alpha \beta} \; ,
\nonumber \\
\sum_\alpha \left (V^*_{\alpha i} V_{\alpha j} \right ) & = & ~
\delta_{i j} \; .
\end{eqnarray}
The six orthogonality relations define six triangles in the
complex plane, as first discussed in Ref. \cite{FXreview}.
Six unitarity triangles have 18 different sides and 9 different
inner angles, but their areas are all identical to
${\cal J}/2$. So far, some particular attention has been paid to
triangles $\triangle_3$ \cite{FXreview} and
$\triangle_\tau$ \cite{Smirnov} shown in Fig. 1. Because current
experimental data indicate that $V$ has a nearly bi-maximal mixing
pattern with $|V_{e3}| \ll 1$, one can easily observe that three sides
of $\triangle_3$ are comparable in magnitude; i.e.,
$|V_{e1} V^*_{e2}| \sim |V_{\mu 1} V^*_{\mu 2}|
\sim |V_{\tau 1} V^*_{\tau 2}| \sim 0.5$. It is therefore possible to
establish $\triangle_3$ and determine its three angles \cite{Xreview},
once its three sides are measured to an acceptable degree of accuracy.
In contrast, one side of
$\triangle_\tau$ is much shorter than its other two sides; i.e.,
$|V_{e1} V^*_{\mu 1}| \sim |V_{e2} V^*_{\mu 2}| \gg |V_{e3} V^*_{\mu 3}|$.
To establish $\triangle_\tau$ needs much more precise data on
its three sides, which must be able to show
$|V_{e1} V^*_{\mu 1}| + |V_{e3} V^*_{\mu 3}| > |V_{e2} V^*_{\mu 2}|$ or
$|V_{e2} V^*_{\mu 2}| + |V_{e3} V^*_{\mu 3}| >
|V_{e1} V^*_{\mu 1}|$ \cite{Smirnov}. Such an accuracy requirement
is practically impossible to be satisfied in the near future \cite{F1}.

To measure ${\cal J}$ and $|V_{\alpha i}|$
in realistic long-baseline experiments of neutrino oscillations, the
terrestrial matter effects must be taken into account \cite{Barger}. The
probabilities of neutrino oscillations in matter can be expressed in the
same form as those in vacuum, however, if we define the {\it effective}
neutrino masses $\tilde{m}_i$ and the {\it effective} lepton flavor
mixing matrix $\tilde{V}$ in which the terrestrial matter effects are
already included \cite{Xing00}. In analogy to the definition of
${\cal J}$ in Eq. (2), the {\it effective} CP-violating parameter
$\tilde{\cal J}$ in matter can be defined as
\begin{equation}
{\rm Im} ( \tilde{V}_{\alpha i} \tilde{V}_{\beta j}
\tilde{V}^*_{\alpha j} \tilde{V}^*_{\beta i} ) =
\tilde{\cal J} \sum_{\gamma, k}
( \epsilon^{~}_{\alpha\beta\gamma} \epsilon^{~}_{ijk} ) \; ,
\end{equation}
where $(\alpha, \beta, \gamma)$ and $(i,j,k)$ run respectively
over $(e, \mu, \tau)$ and $(1,2,3)$. One may similarly define the unitarity
triangles in matter with the help of the unitarity conditions
\begin{eqnarray}
\sum_i \left (\tilde{V}^*_{\alpha i} \tilde{V}_{\beta i} \right )
& = & ~ \delta_{\alpha \beta} \; ,
\nonumber \\
\sum_\alpha \left (\tilde{V}^*_{\alpha i} \tilde{V}_{\alpha j} \right )
& = & ~ \delta_{i j} \; .
\end{eqnarray}
For example, the {\it effective} counterparts of triangles $\triangle_3$
and $\triangle_\tau$ are denoted respectively by
$\tilde{\triangle}_3$ and $\tilde{\triangle}_\tau$.
Because of the terrestrial matter effects, the shapes of $\tilde{\Delta}_3$
and $\tilde{\Delta}_\tau$ are likely to be dramatically different from
those of $\Delta_3$ and $\Delta_\tau$. It is then possible to have
$|\tilde{V}_{e1} \tilde{V}^*_{\mu 1}| \sim
|\tilde{V}_{e2} \tilde{V}^*_{\mu 2}| \sim
|\tilde{V}_{e3} \tilde{V}^*_{\mu 3}|$ for some proper values of the
neutrino beam energy. If this speculation is really true, one will be
able to calculate $\tilde{\cal J}$ by using three sides of
$\tilde{\triangle}_\tau$.

The purpose of this article is to carry out a systematic analysis
of leptonic unitarity triangles in matter. We shall derive the
exact analytical relations between $|V_{\alpha i}|^2$ and
$|\tilde{V}_{\alpha i}|^2$ for a constant matter density profile.
The sides of $\tilde{\triangle}_i$ (for $i=1,2,3$) and
$\tilde{\triangle}_\alpha$ (for $\alpha=e, \mu, \tau$) can then be
linked to those of $\triangle_i$ and $\triangle_\alpha$. The inner
angles of $\tilde{\triangle}_i$ and $\tilde{\triangle}_\alpha$
will also be calculated in terms of the inner angles of
$\triangle_i$ and $\triangle_\alpha$. In addition, we shall
discuss the matter-modified rephasing invariants of $V$ (such as
$\tilde{\cal J}$ and the off-diagonal asymmetries of $\tilde{V}$)
and illustrate the typical shape changes of $\tilde{\triangle}_i$
and $\tilde{\triangle}_\alpha$ with the neutrino beam energy in a
long-baseline experiment. Our results are expected to be very
useful for a complete study of lepton flavor mixing and CP
violation in the era of precision measurements.

The remaining parts of this article are organized as follows. In section II,
we outline the master formulas to derive $\tilde{m}_i$ and $\tilde{V}$
in terms of $m_i$ and $V$. Section III is devoted to the calculation of
$|\tilde{V}_{\alpha i}|^2$. The analytical relations between the inner
angles of $\tilde{\triangle}_i$ (or $\tilde{\triangle}_\alpha$) and
$\triangle_i$ (or $\triangle_\alpha$) are presented in section IV.
Section V is devoted to illustrating the shape evolution
of $\tilde{\triangle}_i$ and $\tilde{\triangle}_\alpha$ with the matter
density. Finally, a brief summary is given in section VI.

\section{Framework}

In the flavor basis chosen in Eq. (1), the
effective Hamiltonians responsible for the propagation of neutrinos in
vacuum and in matter can respectively be written as
\begin{eqnarray}
{\cal H} & = & \frac{1}{2E} \left [ V
\left ( \matrix{
m^2_1 & 0 & 0 \cr
0 & m^2_2 & 0 \cr
0 & 0 & m^2_3 \cr} \right ) V^\dagger \right ] \; ,
\nonumber \\
\tilde{\cal H} & = & \frac{1}{2E} \left [ \tilde{V}
\left ( \matrix{
\tilde{m}^2_1 & 0 & 0 \cr
0 & \tilde{m}^2_2 & 0 \cr
0 & 0 & \tilde{m}^2_3 \cr} \right ) \tilde{V}^\dagger \right ] \; ;
\end{eqnarray}
and their difference
\begin{equation}
\tilde{\cal H} - {\cal H} \; = \; \left ( \matrix{
a & 0 & 0 \cr
0 & 0 & 0 \cr
0 & 0 & 0 \cr} \right ) \;
\end{equation}
signifies the matter effect \cite{MSW}, where $E$ denotes
the neutrino beam energy, $a = \sqrt{2} G_{\rm F} N_e$ measures the
charged-current contribution to the $\nu_e e^-$ forward scattering,
and $N_e$ is the background density of electrons. In writing out
the expression of $\tilde{\cal H}$, we have assumed that the matter
density profile is constant (namely, $N_e =$ constant). Such an
assumption is actually close to reality for most of the proposed
terrestrial long-baseline neutrino oscillation experiments \cite{M}.
In order to establish the analytical relationship between
$\tilde{V}_{\alpha i}$ and $V_{\alpha i}$, we define two quantities
\begin{eqnarray}
p^{~}_{\alpha \beta } & = & 2 E {\cal H}_{\alpha \beta } \; ,
\nonumber \\
q^{~}_{\alpha \beta } & = & \left (2E \right )^2 {\cal H}^{-1}
\det \cal H \;
\end{eqnarray}
in vacuum and their effective counterparts
\begin{eqnarray}
\tilde{p}^{~}_{\alpha \beta } & = & 2 E \tilde{\cal H}_{\alpha \beta } \; ,
\nonumber \\
\tilde{q}^{~}_{\alpha \beta } & = & \left (2E \right )^2 \tilde{\cal H}^{-1}
\det \tilde{\cal H} \;
\end{eqnarray}
in matter \cite{Kimura}, where the subscripts $\alpha$ and $\beta$
run over $e$, $\mu$ and $\tau$. The determinants of $\cal H$ and
$\tilde{\cal H}$ in Eqs. (8) and (9) are simply
$\det {\cal H} = m^2_1 m^2_2 m^2_3/(2E)^3$ and
$\det \tilde{\cal H} = \tilde{m}^2_1 \tilde{m}^2_2 \tilde{m}^2_3/(2E)^3$.
In terms of $m^2_i$ (or $\tilde{m}^2_i$) and $V_{\alpha i}V^*_{\beta i}$
(or $\tilde{V}_{\alpha i}\tilde{V}^*_{\beta i}$), one can obtain
\begin{eqnarray}
p^{~}_{\alpha \beta} & = & \sum_{i=1}^3 \left ( m^2_i
V_{\alpha i}V^*_{\beta i} \right ) \; ,
\nonumber \\
\tilde{p}^{~}_{\alpha \beta} & = & \sum_{i=1}^3 \left ( \tilde{m}^2_i
\tilde{V}_{\alpha i}\tilde{V}^*_{\beta i} \right ) \; ;
\end{eqnarray}
and
\begin{eqnarray}
q^{~}_{\alpha \beta} & = & \frac{1}{2} \sum_{k=1}^3 \left (
m^2_i m^2_j V_{\alpha k}V^*_{\beta k} \epsilon^2_{ijk} \right ) \; ,
\nonumber \\
\tilde{q}^{~}_{\alpha \beta} & = & \frac{1}{2} \sum_{k=1}^3 \left (
\tilde{m}^2_i \tilde{m}^2_j \tilde{V}_{\alpha k}\tilde{V}^*_{\beta k}
\epsilon^2_{ijk} \right ) \; .
\end{eqnarray}
Although the exact analytical relations between
$\tilde{V}_{\alpha i}$ and $V_{\alpha i}$ have been derived in
Ref. \cite{Xing00}, they are not simple enough to calculate
$\tilde{V}_{\alpha i} \tilde{V}^*_{\beta i}$ (or
$\tilde{V}_{\alpha i} \tilde{V}^*_{\alpha j}$), which are directly
relevant to the leptonic unitarity triangles in matter. Hence we shall
establish the relations between
$\tilde{V}_{\alpha i} \tilde{V}^*_{\beta i}$ and
$V_{\alpha i} V^*_{\beta i}$ in a different and simpler way.
Our strategy is as follows: first, we express
$V_{\alpha i} V^*_{\beta i}$ in terms of
$(p^{~}_{\alpha \beta}, q^{~}_{\alpha \beta})$ and
$\tilde{V}_{\alpha i} \tilde{V}^*_{\beta i}$ in terms of
$(\tilde{p}^{~}_{\alpha \beta}, \tilde{q}^{~}_{\alpha \beta})$;
second, we find out the relationship between
$(p^{~}_{\alpha \beta}, q^{~}_{\alpha \beta})$ and
$(\tilde{p}^{~}_{\alpha \beta}, \tilde{q}^{~}_{\alpha \beta})$;
finally, we derive the direct relations between
$\tilde{V}_{\alpha i} \tilde{V}^*_{\beta i}$ and
$V_{\alpha i} V^*_{\beta i}$.

Eqs. (3), (5), (10) and (11) allow us to express
$V_{\alpha i} V^*_{\beta i}$ in terms of
$(p^{~}_{\alpha \beta}, q^{~}_{\alpha \beta})$ and
$\tilde{V}_{\alpha i} \tilde{V}^*_{\beta i}$ in terms of
$(\tilde{p}^{~}_{\alpha \beta}, \tilde{q}^{~}_{\alpha \beta})$.
To see this point more clearly, we introduce the
coefficient matrices $O$ and $\tilde{O}$:
\begin{eqnarray}
O & = &
\left ( \matrix{
1 & 1 & 1 \cr
m_{1}^{2} & m_{2}^{2} & m_{3}^{2} \cr
m_{2}^{2}m_{3}^{2} & m_{1}^{2}m_{3}^{2} & m_{1}^{2}m_{2}^{2} \cr}
\right ) \; ,
\nonumber \\
\tilde{O} & = &
\left ( \matrix{
1 & 1 & 1 \cr
\tilde{m}_{1}^{2} & \tilde{m}_{2}^{2} & \tilde{m}_{3}^{2} \cr
\tilde{m}_{2}^{2}\tilde{m}_{3}^{2} &
\tilde{m}_{1}^{2}\tilde{m}_{3}^{2} &
\tilde{m}_{1}^{2}\tilde{m}_{2}^{2} \cr} \right ) \; .
\end{eqnarray}
Because three neutrino masses are not exactly degenerate,
the inverse matrices of $O$ and $\tilde{O}$ exist. Then one can
obtain
\begin{eqnarray}
\left ( \matrix{
V_{\alpha 1} V_{\beta 1}^{\ast} \cr
V_{\alpha 2} V_{\beta 2}^{\ast} \cr
V_{\alpha 3} V_{\beta 3}^{\ast} \cr} \right )
& = & O^{-1}
\left ( \matrix{
\delta_{\alpha \beta} \cr
p_{\alpha \beta} \cr
q_{\alpha \beta} \cr} \right ) \; ,
\nonumber \\
\left ( \matrix{
\tilde{V}_{\alpha 1} \tilde{V}_{\beta 1}^{\ast} \cr
\tilde{V}_{\alpha 2} \tilde{V}_{\beta 2}^{\ast} \cr
\tilde{V}_{\alpha 3} \tilde{V}_{\beta 3}^{\ast} \cr} \right )
& = & \tilde{O}^{-1}
\left ( \matrix{
\delta_{\alpha \beta} \cr
\tilde{p}_{\alpha \beta} \cr
\tilde{q}_{\alpha \beta} \cr} \right ) \; .
\end{eqnarray}
Taking account of Eqs. (6) and (7), we find that
$(p^{~}_{\alpha \beta}, q^{~}_{\alpha \beta})$
and $(\tilde{p}^{~}_{\alpha \beta}, \tilde{q}^{~}_{\alpha \beta})$
are connected with each other through
\begin{eqnarray}
\tilde{p}^{~}_{\alpha \beta} & = & p^{~}_{\alpha \beta} +
A \left ( \matrix{
1 & ~ 0 ~ & 0 \cr
0 & ~ 0 ~ & 0 \cr
0 & ~ 0 ~ & 0 \cr} \right ) \; ,
\nonumber \\
\tilde{q}^{~}_{\alpha \beta} & = & q^{~}_{\alpha \beta } +
A \left ( \matrix{
0 & 0 & 0 \cr
0 & p^{~}_{\tau \tau} & -p^{~}_{\tau \mu} \cr
0 & -p^{~}_{\mu \tau} & p^{~}_{\mu \mu} \cr} \right ) \; ,
\end{eqnarray}
where $A\equiv 2Ea$. A combination of Eqs. (13) and (14) will
lead to the direct relations between
$\tilde{V}_{\alpha i} \tilde{V}^*_{\beta i}$ and
$V_{\alpha i} V^*_{\beta i}$, as one can see in the subsequent
sections. The usefulness of Eqs. (10)--(14) for discussing the
matter-induced properties of lepton flavor mixing and CP violation
has partly shown up in the literature (see, e.g.,
Refs. \cite{Xreview,Kimura,Xing01,HS}) and will be further
demonstrated in the following.

It is worth mentioning that the generic results obtained above are
only valid for neutrinos propagating in vacuum and in matter. As for
antineutrinos, the corresponding formulas can straightforwardly
be written out from Eqs. (6)--(14) through the replacements
$V\Longrightarrow V^*$ and $A\Longrightarrow -A$.

\section{Moduli ($\alpha = \beta$)}

We first derive the relations between $|\tilde{V}_{\alpha i}|^2$
and $|V_{\alpha i}|^2$ by taking $\alpha = \beta$ for Eq. (13).
Once Eq. (14) is taken into account, we may concretely obtain
\begin{eqnarray}
\left( \matrix{
|\tilde{V}_{e1}|^{2} \cr
|\tilde{V}_{e2}|^{2} \cr
|\tilde{V}_{e3}|^{2} \cr} \right)
& = & \tilde{O}^{-1}O \left( \matrix{
|V_{e1}|^{2} \cr
|V_{e2}|^{2} \cr
|V_{e3}|^{3} \cr} \right)
+ A\tilde{O}^{-1} \left( \matrix{
0 \cr
1 \cr
0 \cr} \right) \; ,
\nonumber \\
\left( \matrix{
|\tilde{V}_{\mu 1}|^{2} \cr
|\tilde{V}_{\mu 2}|^{2} \cr
|\tilde{V}_{\mu 3}|^{2} \cr} \right)
& = & \tilde{O}^{-1}O \left( \matrix{
|V_{\mu 1}|^{2} \cr
|V_{\mu 2}|^{2} \cr
|V_{\mu 3}|^{3} \cr} \right)
+ A\tilde{O}^{-1}T \left( \matrix{
|V_{\tau 1}|^{2} \cr
|V_{\tau 2}|^{2} \cr
|V_{\tau 3}|^{2} \cr}\right) \; ,
\nonumber \\
\left( \matrix{ |\tilde{V}_{\tau 1}|^{2} \cr
|\tilde{V}_{\tau 2}|^{2} \cr
|\tilde{V}_{\tau 3}|^{2} \cr} \right)
& = & \tilde{O}^{-1}O \left( \matrix{
|V_{\tau 1}|^{2} \cr
|V_{\tau 2}|^{2} \cr
|V_{\tau 3}|^{2} \cr} \right)
+ A\tilde{O}^{-1}T \left( \matrix{
|V_{\mu 1}|^{2} \cr
|V_{\mu 2}|^{2} \cr
|V_{\mu 3}|^{3} \cr} \right) \; ,
\end{eqnarray}
where $T$ is defined as
\begin{equation}
T \; = \; \left( \matrix{
0 & 0 & 0 \cr
0 & 0 & 0 \cr
m_{1}^{2} & m_{2}^{2} & m_{3}^{2} \cr} \right) \; .
\end{equation}
In Eq. (15), the inverse matrix of $\tilde{O}$ reads
\begin{equation}
\tilde{O}^{-1} \; =\; \frac{1}{\widetilde{\Delta}_{12}\widetilde{\Delta}_{23}
\widetilde{\Delta}_{31}}
\left( \matrix{
\tilde{m}_{1}^{2}\widetilde{\Delta}_{23}(\tilde{m}_{2}^{2} +
\tilde{m}_{3}^{2})
& \tilde{m}_{1}^{2}\widetilde{\Delta}_{23}
& \widetilde{\Delta}_{23} \cr
\tilde{m}_{2}^{2}\widetilde{\Delta}_{31}(\tilde{m}_{1}^{2} +
\tilde{m}_{3}^{2})
& \tilde{m}_{2}^{2}\widetilde{\Delta}_{31}
& \widetilde{\Delta}_{31} \cr
\tilde{m}_{3}^{2}\widetilde{\Delta}_{12}(\tilde{m}_{1}^{2} +
\tilde{m}_{2}^{2})
& \tilde{m}_{3}^{2}\widetilde{\Delta}_{12}
& \widetilde{\Delta}_{12} \cr} \right) \; ,
\end{equation}
where $\widetilde{\Delta}_{ij} \equiv \tilde{m}^2_i - \tilde{m}^2_j$.
With the help of Eqs. (3) and (5) as well as the relationship \cite{Xreview}
\begin{equation}
\sum_{i=1}^{3}\tilde{m}_{i}^{2} \; = \; \sum_{i=1}^{3}
m_{i}^{2} + A \; ,
\end{equation}
we solve Eq. (15) and arrive at
\begin{eqnarray}
|\tilde{V}_{e1}|^{2} & = & \frac{\Delta_{31}\widehat{\Delta}_{21}}
{\widetilde{\Delta}_{31}\widetilde{\Delta}_{21}}|V_{e1}|^{2} +
\frac{\Delta_{32}\widehat{\Delta}_{11}}{\widetilde{\Delta}_{12}
\widetilde{\Delta}_{13}}|V_{e2}|^{2}+\frac{\widehat{\Delta}_{11}
\widehat{\Delta}_{21}}{\widetilde{\Delta}_{12}\widetilde{\Delta}_{13}} \ ,
\nonumber \\
|\tilde{V}_{e2}|^{2} & = & \frac{\Delta_{12}\widehat{\Delta}_{32}}
{\widetilde{\Delta}_{12}\widetilde{\Delta}_{32}}|V_{e2}|^{2} +
\frac{\Delta_{13}\widehat{\Delta}_{22}}{\widetilde{\Delta}_{21}
\widetilde{\Delta}_{23}}|V_{e3}|^{2}+\frac{\widehat{\Delta}_{22}
\widehat{\Delta}_{32}}{\widetilde{\Delta}_{21}\widetilde{\Delta}_{23}} \ ,
\nonumber \\
|\tilde{V}_{e3}|^{2} & = & \frac{\Delta_{23}\widehat{\Delta}_{13}}
{\widetilde{\Delta}_{23}\widetilde{\Delta}_{13}}|V_{e3}|^{2} +
\frac{\Delta_{21}\widehat{\Delta}_{33}}{\widetilde{\Delta}_{31}
\widetilde{\Delta}_{32}}|V_{e1}|^{2}+\frac{\widehat{\Delta}_{33}
\widehat{\Delta}_{13}}{\widetilde{\Delta}_{31} \widetilde{\Delta}_{32}} \ ;
\end{eqnarray}
and
\begin{eqnarray}
|\tilde{V}_{\mu 1}|^{2} & = & \frac{\Delta_{31}\widehat{\Delta}_{21}}
{\widetilde{\Delta}_{31}\widetilde{\Delta}_{21}}|V_{\mu 1}|^{2} +
\frac{\Delta_{32}\widehat{\Delta}_{11}}{\widetilde{\Delta}_{12}
\widetilde{\Delta}_{13}}|V_{\mu 2}|^{2}+\frac{A\Delta_{13}}
{\widetilde{\Delta}_{12}\widetilde{\Delta}_{13}}|V_{\tau 1}|^{2} +
\frac{A\Delta_{23}}{\widetilde{\Delta}_{12}\widetilde{\Delta}_{13}}
|V_{\tau 2}|^{2} + C_{1} \ ,
\nonumber \\
|\tilde{V}_{\mu 2}|^{2} & = & \frac{\Delta_{12}\widehat{\Delta}_{32}}
{\widetilde{\Delta}_{12}\widetilde{\Delta}_{32}}|V_{\mu 2}|^{2} +
\frac{\Delta_{13}\widehat{\Delta}_{22}}{\widetilde{\Delta}_{21}
\widetilde{\Delta}_{23}}|V_{\mu 3}|^{2}+\frac{A\Delta_{21}}
{\widetilde{\Delta}_{21}\widetilde{\Delta}_{23}}|V_{\tau 2}|^{2} +
\frac{A\Delta_{31}}{\widetilde{\Delta}_{21}\widetilde{\Delta}_{23}}
|V_{\tau 3}|^{2} + C_{2} \ ,
\nonumber \\
|\tilde{V}_{\mu 3}|^{2} & = & \frac{\Delta_{23}\widehat{\Delta}_{13}}
{\widetilde{\Delta}_{23}\widetilde{\Delta}_{13}}|V_{\mu 3}|^{2} +
\frac{\Delta_{21}\widehat{\Delta}_{33}}{\widetilde{\Delta}_{31}
\widetilde{\Delta}_{32}}|V_{\mu 1}|^{2}+\frac{A\Delta_{32}}
{\widetilde{\Delta}_{31}\widetilde{\Delta}_{32}}|V_{\tau 3}|^{2} +
\frac{A\Delta_{12}}{\widetilde{\Delta}_{31}\widetilde{\Delta}_{32}}
|V_{\tau 1}|^{2} + C_{3} \ ;
\end{eqnarray}
and
\begin{eqnarray}
|\tilde{V}_{\tau 1}|^{2} & = & \frac{\Delta_{31}\widehat{\Delta}_{21}}
{\widetilde{\Delta}_{31}\widetilde{\Delta}_{21}}|V_{\tau 1}|^{2} +
\frac{\Delta_{32}\widehat{\Delta}_{11}}{\widetilde{\Delta}_{12}
\widetilde{\Delta}_{13}}|V_{\tau 2}|^{2}+\frac{A\Delta_{13}}
{\widetilde{\Delta}_{12}\widetilde{\Delta}_{13}}|V_{\mu 1}|^{2} +
\frac{A\Delta_{23}}{\widetilde{\Delta}_{12}\widetilde{\Delta}_{13}}
|V_{\mu 2}|^{2} + C_{1} \ ,
\nonumber  \\
|\tilde{V}_{\tau 2}|^{2} & = & \frac{\Delta_{12}\widehat{\Delta}_{32}}
{\widetilde{\Delta}_{12}\widetilde{\Delta}_{32}}|V_{\tau 2}|^{2} +
\frac{\Delta_{13}\widehat{\Delta}_{22}}{\widetilde{\Delta}_{21}
\widetilde{\Delta}_{23}}|V_{\tau 3}|^{2}+\frac{A\Delta_{21}}
{\widetilde{\Delta}_{21}\widetilde{\Delta}_{23}}|V_{\mu 2}|^{2} +
\frac{A\Delta_{31}}{\widetilde{\Delta}_{21}\widetilde{\Delta}_{23}}
|V_{\mu 3}|^{2} + C_{2} \ ,
\nonumber \\
|\tilde{V}_{\tau 3}|^{2} & = &\frac{\Delta_{23}\widehat{\Delta}_{13}}
{\widetilde{\Delta}_{23}\widetilde{\Delta}_{13}}|V_{\tau 3}|^{2} +
\frac{\Delta_{21}\widehat{\Delta}_{33}}{\widetilde{\Delta}_{31}
\widetilde{\Delta}_{32}}|V_{\tau 1}|^{2}+\frac{A\Delta_{32}}
{\widetilde{\Delta}_{31}\widetilde{\Delta}_{32}}|V_{\mu 3}|^{2} +
\frac{A\Delta_{12}}{\widetilde{\Delta}_{31}\widetilde{\Delta}_{32}}
|V_{\mu 1}|^{2} + C_{3} \ ,
\end{eqnarray}
where $\Delta_{ij} \equiv m^2_i - m^2_j$,
$\widehat{\Delta}_{ij} \equiv m^2_i - \tilde{m}^2_j$, and
\begin{eqnarray}
C_{1} & = & -\frac{\Delta_{31}\Delta_{23}+\widehat{\Delta}_{31}
\widehat{\Delta}_{32}+\widehat{\Delta}_{31}\widehat{\Delta}_{33}}
{\widetilde{\Delta}_{12}\widetilde{\Delta}_{13}} \ ,
\nonumber \\
C_{2} & = & -\frac{\Delta_{12}\Delta_{31}+\widehat{\Delta}_{12}
\widehat{\Delta}_{13}+\widehat{\Delta}_{12}\widehat{\Delta}_{11}}
{\widetilde{\Delta}_{21}\widetilde{\Delta}_{23}} \ ,
\nonumber \\
C_{3} & = & -\frac{\Delta_{23}\Delta_{12}+\widehat{\Delta}_{23}
\widehat{\Delta}_{21}+\widehat{\Delta}_{23}\widehat{\Delta}_{22}}
{\widetilde{\Delta}_{31}\widetilde{\Delta}_{32}} \ .
\end{eqnarray}
In appendix A, we list the explicit expressions of
$\widetilde{\Delta}_{ij}$ and $\widehat{\Delta}_{ij}$ in terms of
$\Delta_{ij}$ and $A$. It is then possible to evaluate the deviation of
$|\tilde{V}_{\alpha i}|^2$ from $|V_{\alpha i}|^2$ by using
the formulas obtained above. Of course,
$|\tilde{V}_{\alpha i}|^2 = |V_{\alpha i}|^2$ holds at the
limit $A \rightarrow 0$ or $E \rightarrow 0$.

Two sets of normalization conditions given in Eq. (5) allow us to
define two off-diagonal asymmetries of $\tilde{V}$:
\begin{eqnarray}
\tilde{\cal A}_{\rm L} & \equiv &
|\tilde{V}_{e2}|^{2}-|\tilde{V}_{\mu 1}|^{2} =
|\tilde{V}_{\mu 3}|^{2}-|\tilde{V}_{\tau 2}|^{2} =
|\tilde{V}_{\tau 1}|^{2}-|\tilde{V}_{e3}|^{2} \ ,
\nonumber \\
\tilde{\cal A}_{\rm R} & \equiv &
|\tilde{V}_{e2}|^{2}-|\tilde{V}_{\mu 3}|^{2} =
|\tilde{V}_{\mu 1}|^{2}-|\tilde{V}_{\tau 2}|^{2} =
|\tilde{V}_{\tau 3}|^{2}-|\tilde{V}_{e1}|^{2} \ .
\end{eqnarray}
They can straightforwardly be calculated by using Eqs. (19), (20) and (21).
Comparing between $\tilde{\cal A}_{\rm L}$ (or $\tilde{\cal A}_{\rm R}$)
and its counterpart ${\cal A}_{\rm L}$ (or ${\cal A}_{\rm R}$) in vacuum,
we may examine the matter effect on the geometric structure
of $V$. In particular, the six matter-modified unitarity triangles will
reduce to three pairs of {\it congruent} triangles \cite{Xing02}, if
$\tilde{\cal A}_{\rm L}=0$ or $\tilde{\cal A}_{\rm R} = 0$ holds.

The sides of six unitarity triangles can also be
obtained from Eqs. (19), (20) and (21). If three sides of a specific
triangle are comparable in magnitude, one may use them to calculate
the area of this triangle -- namely, the CP-violating invariant
$\tilde{\cal J}$. Indeed, $\tilde{\cal J}$ is given by
\begin{eqnarray}
\tilde{\cal J}^2 & = & |\tilde{V}_{\alpha i}|^2 |\tilde{V}_{\beta j}|^2
|\tilde{V}_{\alpha j}|^2 |\tilde{V}_{\beta i} |^2 -
\frac{1}{4} \left (1 + |\tilde{V}_{\alpha i}|^2 |\tilde{V}_{\beta j}|^2
+ |\tilde{V}_{\alpha j}|^2 |\tilde{V}_{\beta i}|^2 \right .
\nonumber \\
& & \left . - |\tilde{V}_{\alpha i}|^2 - |\tilde{V}_{\beta j}|^2
- |\tilde{V}_{\alpha j}|^2 - |\tilde{V}_{\beta i}|^2 \right )^2 \; ,
\end{eqnarray}
in which $\alpha \neq \beta$ running over $(e, \mu, \tau)$ and
$i \neq j$ running over $(1, 2, 3)$. The implication of this result is
obvious: important information about leptonic CP violation can in
principle be extracted from the measured moduli of four independent
matrix elements of $\tilde{V}$.

\section{Angles ($\alpha \neq \beta$)}

Now let us focus our attention on the inner angles of six unitarity
triangles. There are only 9 independent angles, which appear either
in triangles $\triangle_{e,\mu,\tau}$ or in triangles $\triangle_{1,2,3}$
(see Fig. 1 for illustration). Hence it is only necessary to consider
$(\triangle_e, \triangle_\mu, \triangle_\tau)$ and their effective
counterparts
$(\tilde{\triangle}_e, \tilde{\triangle}_\mu, \tilde{\triangle}_\tau)$
in the calculation of 9 angles. To be specific, an inner angle of
$\Delta_\alpha$ or $\tilde{\Delta}_\alpha$ (for $\alpha = e, \mu$ or
$\tau$) can be defined as
\begin{eqnarray}
\phi^{ij}_{\alpha\beta} & \equiv & \arg \left (
- \frac{V_{\alpha i}V^*_{\beta i}}{V_{\alpha j}V^*_{\beta j}} \right ) \; ,
\nonumber \\
\tilde{\phi}^{ij}_{\alpha\beta} & \equiv & \arg \left (
- \frac{\tilde{V}_{\alpha i}\tilde{V}^*_{\beta i}}{\tilde{V}_{\alpha j}
\tilde{V}^*_{\beta j}} \right ) \; ,
\end{eqnarray}
where $(\alpha,\beta)$ run over $(e,\mu)$, $(\mu ,\tau)$ and
$(\tau ,e)$, and $(i,j)$ run over $(1,2)$, $(2,3)$ and $(3,1)$. Taking
account of Eqs. (2) and (4), we obtain
\begin{eqnarray}
\cot\phi^{ij}_{\alpha\beta} & = & \frac{1}{\cal J}
{\rm Re}\left (V_{\alpha i} V_{\beta j} V^*_{\alpha j}
V^*_{\beta i} \right ) \; ,
\nonumber \\
\cot\tilde{\phi}^{ij}_{\alpha\beta} & = & \frac{1}{\tilde{\cal J}}
{\rm Re}\left (\tilde{V}_{\alpha i} \tilde{V}_{\beta j}
\tilde{V}^*_{\alpha j} \tilde{V}^*_{\beta i} \right ) \; .
\end{eqnarray}
Note that $\cal J$ and $\tilde{\cal J}$ are related with each other
through the equation \cite{Naumov}
\begin{equation}
\tilde{\cal J} \widetilde{\Delta}_{12}\widetilde{\Delta}_{13}
\widetilde{\Delta}_{23} \; = \; {\cal J}
\Delta_{12}\Delta_{13}\Delta_{23} \;\; . ~~
\end{equation}
In order to link $\cot\tilde{\phi}^{ij}_{\alpha\beta}$ to
$\cot\phi^{ij}_{\alpha\beta}$ via Eq. (26), we need to find
out the relationship between ${\rm Re}(\tilde{V}_{\alpha i}
\tilde{V}_{\beta j} \tilde{V}^*_{\alpha j} \tilde{V}^*_{\beta i})$
and ${\rm Re} (V_{\alpha i} V_{\beta j} V^*_{\alpha j} V^*_{\beta i})$.

For $\alpha \neq \beta$, Eqs. (13) and (14) yield
\begin{eqnarray}
\left( \matrix{
\tilde{V}_{e1}\tilde{V}_{\mu 1}^{\ast} \cr
\tilde{V}_{e2}\tilde{V}_{\mu 2}^{\ast} \cr
\tilde{V}_{e3}\tilde{V}_{\mu 3}^{\ast} \cr} \right)
& = & \tilde{O}^{-1}O \left( \matrix{
V_{e1}V_{\mu 1}^{\ast} \cr
V_{e2}V_{\mu 2}^{\ast} \cr
V_{e3}V_{\mu 3}^{\ast} \cr} \right) \ ,
\nonumber \\
\left( \matrix{ \tilde{V}_{e1}\tilde{V}_{\tau 1}^{\ast} \cr
\tilde{V}_{e2}\tilde{V}_{\tau 2}^{\ast} \cr
\tilde{V}_{e3}\tilde{V}_{\tau 3}^{\ast} \cr} \right) & = &
\tilde{O}^{-1}O \left( \matrix{ V_{e1}V_{\tau 1}^{\ast} \cr
V_{e2}V_{\tau 2}^{\ast} \cr V_{e3}V_{\tau 3}^{\ast} \cr} \right) \
,
\nonumber \\
\left( \matrix{ \tilde{V}_{\mu 1}\tilde{V}_{\tau 1}^{\ast} \cr
\tilde{V}_{\mu 2}\tilde{V}_{\tau 2}^{\ast} \cr \tilde{V}_{\mu
3}\tilde{V}_{\tau 3}^{\ast} \cr} \right) & = & \tilde{O}^{-1}O
\left( \matrix{ V_{\mu 1}V_{\tau 1}^{\ast} \cr V_{\mu 2}V_{\tau
2}^{\ast} \cr V_{\mu 3}V_{\tau 3}^{\ast} \cr} \right) -
A\tilde{O}^{-1}T \left( \matrix{ V_{\mu 1}V_{\tau 1}^{\ast} \cr
V_{\mu 2}V_{\tau 2}^{\ast} \cr V_{\mu 3}V_{\tau 3}^{\ast} \cr}
\right) \ ,
\end{eqnarray}
where $T$ has been defined in Eq. (16). With the help of
Eqs. (17) and (18), one may solve Eq. (28) and obtain the
matter-induced corrections to $V_{\alpha i}V^*_{\beta i}$.
Three sides of the effective unitarity triangle
$\tilde{\triangle}_{e}$, $\tilde{\triangle}_{\mu}$ or
$\tilde{\triangle}_{\tau}$ are then given by
\begin{eqnarray}
\tilde{V}_{e1}\tilde{V}_{\mu 1}^{\ast}
& = & \frac{\widehat{\Delta}_{21}\Delta_{31}}{\widetilde{\Delta}_{21}
\widetilde{\Delta}_{31}}V_{e1}V_{\mu 1}^{\ast}+\frac{\widehat{\Delta}_{11}
\Delta_{32}}{\widetilde{\Delta}_{12}\widetilde{\Delta}_{13}}
V_{e2}V_{\mu 2}^{\ast} \ ,
\nonumber  \\
\tilde{V}_{e2}\tilde{V}_{\mu 2}^{\ast}
& = & \frac{\widehat{\Delta}_{32}\Delta_{21}}{\widetilde{\Delta}_{32}
\widetilde{\Delta}_{21}}V_{e2}V_{\mu 2}^{\ast}+\frac{\widehat{\Delta}_{22}
\Delta_{31}}{\widetilde{\Delta}_{12}\widetilde{\Delta}_{23}}
V_{e3}V_{\mu 3}^{\ast} \ ,
\nonumber \\
\tilde{V}_{e3}\tilde{V}_{\mu 3}^{\ast}
& = & \frac{\widehat{\Delta}_{13}\Delta_{23}}{\widetilde{\Delta}_{13}
\widetilde{\Delta}_{23}}V_{e3}V_{\mu 3}^{\ast}+\frac{\widehat{\Delta}_{33}
\Delta_{21}}{\widetilde{\Delta}_{13}\widetilde{\Delta}_{23}}
V_{e1}V_{\mu 1}^{\ast}
\end{eqnarray}
for $\tilde{\triangle}_{\tau}$; and
\begin{eqnarray}
\tilde{V}_{\tau 1}\tilde{V}_{e 1}^{\ast}
& = & \frac{\widehat{\Delta}_{21}\Delta_{31}}{\widetilde{\Delta}_{21}
\widetilde{\Delta}_{31}}V_{\tau 1}V_{e 1}^{\ast}+\frac{\widehat{\Delta}_{11}
\Delta_{32}}{\widetilde{\Delta}_{12}\widetilde{\Delta}_{13}}
V_{\tau 2}V_{e 2}^{\ast} \ ,
\nonumber  \\
\tilde{V}_{\tau 2}\tilde{V}_{e 2}^{\ast}
& = & \frac{\widehat{\Delta}_{32}\Delta_{21}}{\widetilde{\Delta}_{32}
\widetilde{\Delta}_{21}}V_{\tau 2}V_{e 2}^{\ast}+\frac{\widehat{\Delta}_{22}
\Delta_{31}}{\widetilde{\Delta}_{12}\widetilde{\Delta}_{23}}
V_{\tau 3}V_{e 3}^{\ast} \ ,
\nonumber  \\
\tilde{V}_{\tau 3}\tilde{V}_{e 3}^{\ast}
& = & \frac{\widehat{\Delta}_{13}\Delta_{23}}{\widetilde{\Delta}_{13}
\widetilde{\Delta}_{23}}V_{\tau 3}V_{e 3}^{\ast}+\frac{\widehat{\Delta}_{33}
\Delta_{21}}{\widetilde{\Delta}_{13}\widetilde{\Delta}_{23}}
V_{\tau 1}V_{e 1}^{\ast}
\end{eqnarray}
for $\tilde{\triangle}_{\mu}$; and
\begin{eqnarray}
\tilde{V}_{\mu 1}\tilde{V}_{\tau 1}^{\ast}
& = & \frac{(\widehat{\Delta}_{21}+A)\Delta_{31}}{\widetilde{\Delta}_{21}
\widetilde{\Delta}_{31}}V_{\mu 1}V_{\tau 1}^{\ast}+
\frac{(\widehat{\Delta}_{11}+A)\Delta_{32}}{\widetilde{\Delta}_{12}
\widetilde{\Delta}_{13}}V_{\mu 2}V_{\tau 2}^{\ast} \ ,
\nonumber  \\
\tilde{V}_{\mu 2}\tilde{V}_{\tau 2}^{\ast}
& = & \frac{(\widehat{\Delta}_{32}+A)\Delta_{21}}{\widetilde{\Delta}_{32}
\widetilde{\Delta}_{21}}V_{\mu 2}V_{\tau 2}^{\ast}+
\frac{(\widehat{\Delta}_{22}+A)\Delta_{31}}{\widetilde{\Delta}_{12}
\widetilde{\Delta}_{23}}V_{\mu 3}V_{\tau 3}^{\ast} \ ,
\nonumber  \\
\tilde{V}_{\mu 3}\tilde{V}_{\tau 3}^{\ast}
& = & \frac{(\widehat{\Delta}_{13}+A)\Delta_{23}}{\widetilde{\Delta}_{13}
\widetilde{\Delta}_{23}}V_{\mu 3}V_{\tau 3}^{\ast}+
\frac{(\widehat{\Delta}_{33}+A)\Delta_{21}}{\widetilde{\Delta}_{13}
\widetilde{\Delta}_{23}} V_{\mu 1}V_{\tau 1}^{\ast}
\end{eqnarray}
for $\tilde{\triangle}_e$. It is remarkable that Eq. (29), (30) or (31),
together with Eqs. (2) and (4), can be used to derive Eq. (27).
The relationship between ${\rm Re}(\tilde{V}_{\alpha i}
\tilde{V}_{\beta j} \tilde{V}^*_{\alpha j} \tilde{V}^*_{\beta i})$
and ${\rm Re} (V_{\alpha i} V_{\beta j} V^*_{\alpha j} V^*_{\beta i})$
can also be derived from these equations. Then we are able to
establish the direct connection between
$\cot\tilde{\phi}^{ij}_{\alpha\beta}$ and $\cot\phi^{ij}_{\alpha\beta}$.

Comparing between the definition of $\phi^{ij}_{\alpha\beta}$ and the
simpler notation of nine inner angles in Fig. 1, we have
$\angle 1 = \phi^{12}_{\mu\tau}$,
$\angle 2 = \phi^{23}_{\mu\tau}$,
$\angle 3 = \phi^{31}_{\mu\tau}$;
$\angle 4 = \phi^{12}_{\tau e}$,
$\angle 5 = \phi^{23}_{\tau e}$,
$\angle 6 = \phi^{31}_{\tau e}$;
$\angle 7 = \phi^{12}_{e \mu}$,
$\angle 8 = \phi^{23}_{e \mu}$,
$\angle 9 = \phi^{31}_{e \mu}$.
One may use the similar notation ($\widetilde{\angle} 1$,
$\cdot\cdot\cdot$, $\widetilde{\angle} 9$) to replace
$\tilde{\phi}^{ij}_{\alpha\beta}$ for the matter-modified unitarity
triangles. Therefore,
\begin{eqnarray}
\cot \widetilde{\angle} 1 & = &
\frac{(\widehat{\Delta}_{32}+A)(\widehat{\Delta}_{21}+A)}
{\Delta_{32}\widetilde{\Delta}_{21}}\cot\angle 1 +
\frac{(\widehat{\Delta}_{11}+A)(\widehat{\Delta}_{22}+A)}
{\Delta_{12}\widetilde{\Delta}_{12}}\cot\angle 2
\nonumber \\
&& + \frac{(\widehat{\Delta}_{21}+A)(\widehat{\Delta}_{22}+A)\Delta_{13}}
{\Delta_{12}\Delta_{23}\widetilde{\Delta}_{12}}\cot\angle 3 +
\frac{(\widehat{\Delta}_{11}+A)(\widehat{\Delta}_{32}+A)}
{{\cal J} \Delta_{31}\widetilde{\Delta}_{12}}
|V_{\mu 2}V_{\tau 2}^{*}|^{2} \ ,
\nonumber \\
\cot \widetilde{\angle} 2 & = &
\frac{(\widehat{\Delta}_{13}+A)(\widehat{\Delta}_{32}+A)}
{\Delta_{13}\widetilde{\Delta}_{32}}\cot\angle 2 +
\frac{(\widehat{\Delta}_{22}+A)(\widehat{\Delta}_{33}+A)}
{\Delta_{23}\widetilde{\Delta}_{23}}\cot\angle 3
\nonumber \\
&& + \frac{(\widehat{\Delta}_{32}+A)(\widehat{\Delta}_{33}+A)\Delta_{21}}
{\Delta_{23}\Delta_{31}\widetilde{\Delta}_{23}}\cot\angle 1 +
\frac{(\widehat{\Delta}_{22}+A)(\widehat{\Delta}_{13}+A)}
{{\cal J}\Delta_{12}\widetilde{\Delta}_{23}}
|V_{\mu 3}V_{\tau 3}^{*}|^{2} \ ,
\nonumber \\
\cot \widetilde{\angle} 3 & = &
\frac{(\widehat{\Delta}_{21}+A)(\widehat{\Delta}_{13}+A)}
{\Delta_{21}\widetilde{\Delta}_{13}}\cot\angle 3 +
\frac{(\widehat{\Delta}_{11}+A)(\widehat{\Delta}_{33}+A)}
{\Delta_{31}\widetilde{\Delta}_{31}}\cot\angle 1
\nonumber \\
&& + \frac{(\widehat{\Delta}_{11}+A)(\widehat{\Delta}_{13}+A)\Delta_{32}}
{\Delta_{12}\Delta_{31}\widetilde{\Delta}_{31}}\cot\angle 2 +
\frac{(\widehat{\Delta}_{21}+A)(\widehat{\Delta}_{33}+A)}
{{\cal J}\Delta_{23}\widetilde{\Delta}_{31}}
|V_{\mu 1}V_{\tau 1}^{*}|^{2} \; ;
\end{eqnarray}
and
\begin{eqnarray}
\cot \widetilde{\angle} 4 & = &
\frac{\widehat{\Delta}_{32}\widehat{\Delta}_{21}}
{\Delta_{32}\widetilde{\Delta}_{21}}\cot\angle 4 +
\frac{\widehat{\Delta}_{11}\widehat{\Delta}_{22}}
{\Delta_{12}\widetilde{\Delta}_{12}}\cot\angle 5 +
\frac{\widehat{\Delta}_{21}\widehat{\Delta}_{22}\Delta_{13}}
{\Delta_{12}\Delta_{23}\widetilde{\Delta}_{12}}\cot\angle 6 +
\frac{\widehat{\Delta}_{11}\widehat{\Delta}_{32}}
{{\cal J}\Delta_{31}\widetilde{\Delta}_{12}}
|V_{\tau 2}V_{e2}^{*}|^{2} \ ,
\nonumber \\
\cot \widetilde{\angle} 5 & = &
\frac{\widehat{\Delta}_{13}\widehat{\Delta}_{32}}
{\Delta_{13}\widetilde{\Delta}_{32}}\cot\angle 5 +
\frac{\widehat{\Delta}_{22}\widehat{\Delta}_{33}}
{\Delta_{23}\widetilde{\Delta}_{23}}\cot\angle 6 +
\frac{\widehat{\Delta}_{32}\widehat{\Delta}_{33}\Delta_{21}}
{\Delta_{23}\Delta_{31}\widetilde{\Delta}_{23}}\cot\angle 4 +
\frac{\widehat{\Delta}_{22}\widehat{\Delta}_{13}}
{{\cal J}\Delta_{12}\widetilde{\Delta}_{23}}
|V_{\tau 3}V_{e3}^{*}|^{2} \ ,
\nonumber \\
\cot \widetilde{\angle} 6 & = &
\frac{\widehat{\Delta}_{21}\widehat{\Delta}_{13}}
{\Delta_{21}\widetilde{\Delta}_{13}}\cot\angle 6 +
\frac{\widehat{\Delta}_{11}\widehat{\Delta}_{33}}
{\Delta_{31}\widetilde{\Delta}_{31}}\cot\angle 4 +
\frac{\widehat{\Delta}_{11}\widehat{\Delta}_{13}\Delta_{32}}
{\Delta_{12}\Delta_{31}\widetilde{\Delta}_{31}}\cot\angle 5 +
\frac{\widehat{\Delta}_{21}\widehat{\Delta}_{33}}
{{\cal J}\Delta_{23}\widetilde{\Delta}_{31}}
|V_{\tau 1}V_{e1}^{*}|^{2} \ ;
\end{eqnarray}
and
\begin{eqnarray}
\cot \widetilde{\angle} 7 & = &
\frac{\widehat{\Delta}_{32}\widehat{\Delta}_{21}}
{\Delta_{32}\widetilde{\Delta}_{21}}\cot\angle 7 +
\frac{\widehat{\Delta}_{11}\widehat{\Delta}_{22}}
{\Delta_{12}\widetilde{\Delta}_{12}}\cot\angle 8 +
\frac{\widehat{\Delta}_{21}\widehat{\Delta}_{22}\Delta_{13}}
{\Delta_{12}\Delta_{23}\widetilde{\Delta}_{12}}\cot\angle 9 +
\frac{\widehat{\Delta}_{11}\widehat{\Delta}_{32}}
{{\cal J}\Delta_{31}\widetilde{\Delta}_{12}}
|V_{e2}V_{\mu 2}^{*}|^{2} \ ,
\nonumber \\
\cot \widetilde{\angle} 8 & = &
\frac{\widehat{\Delta}_{13}\widehat{\Delta}_{32}}
{\Delta_{13}\widetilde{\Delta}_{32}}\cot\angle 8 +
\frac{\widehat{\Delta}_{22}\widehat{\Delta}_{33}}
{\Delta_{23}\widetilde{\Delta}_{23}}\cot\angle 9 +
\frac{\widehat{\Delta}_{32}\widehat{\Delta}_{33}\Delta_{21}}
{\Delta_{23}\Delta_{31}\widetilde{\Delta}_{23}}\cot\angle 7 +
\frac{\widehat{\Delta}_{22}\widehat{\Delta}_{13}}
{{\cal J}\Delta_{12}\widetilde{\Delta}_{23}}
|V_{e3}V_{\mu 3}^{*}|^{2} \ ,
\nonumber \\
\cot \widetilde{\angle} 9 & = &
\frac{\widehat{\Delta}_{21}\widehat{\Delta}_{13}}
{\Delta_{21}\widetilde{\Delta}_{13}}\cot\angle 9 +
\frac{\widehat{\Delta}_{11}\widehat{\Delta}_{33}}
{\Delta_{31}\widetilde{\Delta}_{31}}\cot\angle 7 +
\frac{\widehat{\Delta}_{11}\widehat{\Delta}_{13}\Delta_{32}}
{\Delta_{12}\Delta_{31}\widetilde{\Delta}_{31}}\cot\angle 8 +
\frac{\widehat{\Delta}_{21}\widehat{\Delta}_{33}}
{{\cal J}\Delta_{23}\widetilde{\Delta}_{31}}
|V_{e1}V_{\mu 1}^{*}|^{2} \ .
\end{eqnarray}
These exact analytical results clearly show how nine inner
angles of six unitarity triangles get modified by the matter
effects.

\section{Illustration}

We proceed to numerically illustrate the matter effects on the
shapes of six unitarity triangles, the rephasing invariant of CP
violation, and the off-diagonal asymmetries of $V$. In view of
current solar \cite{SNO} and atmospheric \cite{SK} neutrino
oscillation data, we typically take $\Delta_{21} \approx 8 \times
10^{-5} ~ {\rm eV}^2$ and $\Delta_{32} \approx 2.3 \times 10^{-3}
~ {\rm eV}^2$. We also take $\theta_{12} \approx 33^\circ$,
$\theta_{23} \approx 45^\circ$, $\theta_{13} \approx 3^\circ$ and
$\delta \approx 90^\circ$ in the standard parametrization of $V$
\cite{Xreview}. It is unnecessary to specify two Majorana-type
CP-violating phases of $V$ in our calculations, because they play
no role in the configuration of leptonic unitarity triangles
\cite{Branco}. The inputs taken above lead to ${\cal J} \approx
0.012$, ${\cal A}_{\rm L} \approx 0.147$ and ${\cal A}_{\rm R}
\approx -0.203$. This means that $V$ is asymmetric both about its
$V_{e1}$-$V_{\mu 2}$-$V_{\tau 3}$ axis and about its
$V_{e3}$-$V_{\mu 2}$-$V_{\tau 1}$ axis, and the area of each
triangle is about 8 times smaller than its maximal limit (i.e.,
${\cal J} = 1/(6\sqrt{3})$ \cite{Jarlskog}).

For a realistic long-baseline neutrino oscillation experiment, the
dependence of terrestrial matter effects on the neutrino beam
energy can approximately be written as $A \approx 2.28 \times
10^{-4} {\rm eV}^2 E/[{\rm GeV}]$ \cite{M}. This approximation is
reasonably good and close to reality, only if the baseline length
is about 1000 km or shorter \cite{MM}. Of course, $\tilde{V} = V$
holds in the limit of $E=0$ or $A=0$. Typically taking $E=1$ GeV,
2 GeV and 3 GeV, we calculate the sides of six effective unitarity
triangles in matter and show the changes of their shapes in Figs.
2--5. The corresponding results for 9 inner angles of
$\tilde{\triangle}_{e,\mu,\tau}$ or $\tilde{\triangle}_{1,2,3}$
are presented in Table 1, and the numerical dependence of
$\tilde{\cal J}$, $\tilde{\cal A}_{\rm L}$ and $\tilde{\cal
A}_{\rm R}$ on $E$ is illustrated in Fig. 6. Some comments and
discussions are in order.

(a) The smallness of $\theta_{13}$ (or $|V_{e3}|$) makes one side of
$\triangle_1$ or $\triangle_2$ strongly suppressed. When
the terrestrial matter effect with $E \geq 1$ GeV is taken into account,
three sides of $\tilde{\triangle}_1$ and $\tilde{\triangle}_3$ become
comparable in magnitude for neutrinos ($+A$ and $V$); or
three sides of $\tilde{\triangle}_2$ and $\tilde{\triangle}_3$ become
comparable in magnitude for antineutrinos ($-A$ and $V^*$).
As a consequence of matter
corrections, the shortest side of $\triangle_2$
(i.e., $S^{31}_{ee} \equiv |\tilde{V}_{e3} \tilde{V}^*_{e1}|$ at
$A=0$) turns out to be shorter for neutrinos; so does that of
$\triangle_1$ (i.e.,
$S^{23}_{ee} \equiv |\tilde{V}_{e2} \tilde{V}^*_{e3}|$ at $A=0$)
for antineutrinos. We find that the shape of $\tilde{\triangle}_3$ is
relatively stable against matter corrections, no matter whether the
neutrino beam or the antineutrino beam is concerned.

(b) Similarly because of the smallness of $\theta_{13}$ (or
$|V_{e3}|$), one side of $\triangle_\mu$ or $\triangle_\tau$ is
strongly suppressed. The terrestrial matter effect becomes
significant for $E \geq 1$ GeV. In this case, three sides of
$\tilde{\triangle}_\mu$ and $\tilde{\triangle}_\tau$ are
comparable in magnitude for either neutrinos ($+A$ and $V$) or
antineutrinos ($-A$ and $V^*$). An interesting feature of
$\tilde{\triangle}_e$ is that its side $S^{22}_{\mu\tau} \equiv
|\tilde{V}_{\mu 2} \tilde{V}^*_{\tau 2}|$ is dramatically
sensitive to matter corrections and becomes very short for
neutrinos, while its side $S^{11}_{\mu\tau} \equiv |\tilde{V}_{\mu
1} \tilde{V}^*_{\tau 1}|$ may significantly be suppressed by
matter effects for antineutrinos. One can see that the shapes of
$\tilde{\triangle}_\mu$ and $\tilde{\triangle}_\tau$ are
relatively stable against matter corrections, no matter whether
the neutrino beam or the antineutrino beam is concerned.

(c) Note that the input $\theta_{23} \approx 45^\circ$, which is
well favored by current atmospheric neutrino oscillation data \cite{SK},
results in $|V_{\mu i}| \approx |V_{\tau i}|$ (for $i=1,2,3$). Hence
triangles $\triangle_\mu$ and $\triangle_\tau$ are congruent with
each other; so are their effective counterparts
$\tilde{\triangle}_\mu$ and $\tilde{\triangle}_\tau$. This
accidental result has actually shown up in Figs. 4 and 5. From the
phenomenological point of view, it makes sense to measure $\theta_{23}$
accurately and to examine its possible deviation from maximal mixing
(i.e., $\theta_{23} = 45^\circ$ exactly),
so as to explore the underlying $\mu$-$\tau$ flavor
symmetry and its geometric manifestation in leptonic unitarity triangles.

(d) Table 1 is helpful for us to understand the terrestrial
matter effect on 9 inner angles of six unitarity triangles.
One can see that $\tilde{\angle} 6$ and $\tilde{\angle} 9$, which
happen to be identical as a consequence of $\theta_{23} \approx 45^\circ$,
are relatively stable against matter corrections. In contrast,
$\tilde{\angle} 2$, $\tilde{\angle} 3$ and $\tilde{\angle} 4$ are
rather sensitive to matter corrections.

(e) No matter how the sides and angles of six unitarity triangles
change with the matter effect, their area ($\tilde{\cal J}/2$) is in most
cases smaller than that in vacuum (${\cal J}/2$). This unfortunate
feature, as shown in Fig. 6, makes it hard to directly measure leptonic
CP or T violation in any realistic long-baseline neutrino oscillations. If
the neutrino beam energy is small (e.g., about 1 GeV or smaller), the
matter-induced suppression of ${\cal J}$ is not significant. In this
case, a study of leptonic CP violation and unitarity triangles in a
{\it medium}-baseline neutrino oscillation experiment seems more
feasible \cite{MLBL}.

(f) The matter effect on two off-diagonal asymmetries of $V$ is also
illustrated in Fig. 6. One can see that
$\tilde{\cal A}_{\rm L} \approx 0$ is a good approximation, when the
beam energy of antineutrinos satisfies $E \geq 1$ GeV. In this interesting
case, one may approximately arrive at
$\tilde{\triangle}_e \cong \tilde{\triangle}_1$,
$\tilde{\triangle}_\mu \cong \tilde{\triangle}_2$ and
$\tilde{\triangle}_\tau \cong \tilde{\triangle}_3$. Such a result can
also be seen from Figs. 3 and 5.
The possibility for $\tilde{\cal A}_{\rm R} \approx 0$ may appear only
when the beam energy of neutrinos is around $E \approx 0.2$ GeV, as
shown in Fig. 6. These results are certainly dependent upon our inputs.
Therefore, they mainly serve for illustration.

In practice, the reconstruction of a unitarity triangle requires a
series of measurements, which can be done in both {\it appearance}
and {\it disappearance} neutrino oscillation experiments. More detailed
analyses in this respect can be seen from Ref. \cite{Smirnov} (see
also Ref. \cite{Sato}), where the central attention has only been paid
to the triangle $\triangle_\tau$ and its effective counterpart
$\tilde{\triangle}_\tau$. A similar analysis of other triangles is
expected to be very lengthy and will be presented elsewhere \cite{XZ05}.
As we have emphasized, it is possible to establish leptonic CP violation
by means of the precise experimental information about the sides of
six unitarity triangles. In this sense, the geometric approach
described here may be complimentary to the direct determination of
CP violation from the measurement of probability asymmetries between
$\nu_\alpha \rightarrow \nu^{~}_\beta$ and
$\overline{\nu}_\alpha \rightarrow \overline{\nu}^{~}_\beta$
(for $\alpha \neq \beta$ running over $e$, $\mu$, $\tau$) oscillations.

\section{Summary}

Considering the lepton flavor mixing matrix $V$ and its effective
counterpart $\tilde{V}$ in matter, we have carried out a systematic
analysis of their corresponding unitarity triangles in the complex
plane. The exact analytical relations between the moduli
$|V_{\alpha i}|$ and $|\tilde{V}_{\alpha i}|$ have been derived. The sides
and angles of each unitarity triangle have also been calculated by taking
account of the terrestrial matter effect. We have illustrated the
shape evolution of six effective triangles with the beam energy of
neutrinos and antineutrinos in a realistic long-baseline oscillation
experiment. Matter corrections to the rephasing-invariant parameter
of CP violation and the off-diagonal asymmetries of $V$ have been
discussed too.

We expect that this complete geometric description of
matter-modified lepton flavor mixing and CP violation will be very
useful for the future long-baseline neutrino oscillation experiments,
although it remains unclear how far our experimentalists will go in
this direction. We admit that how to experimentally realize the proposed
ideas and methods is certainly a challenging question.
However, it is absolutely clear that the measurement
of leptonic CP violation must be a top task of experimental neutrino
physics in the coming years or even decades. Let us recall what has
happened in the quark sector: CP violation and one unitarity triangle
(defined by $V_{ud}V^*_{ub} + V_{cd}V^*_{cb} + V_{td}V^*_{tb} = 0$
in the complex plane) have been established at KEK-$B$ and SLAC-$B$
factories, and a further study of other five triangles will be
implemented at LHC-$B$ in the near future. Thus we are convinced that
similar steps will be taken for an experimental determination of
leptonic CP violation and unitarity triangles in the era of
{\it precision} neutrino physics.

\vspace{0.5cm}

This work was supported in part by the National Nature Science
Foundation of China.

\newpage

\appendix
\section{}

In the assumption of a constant earth density profile and with the
help of the effective Hamiltonians given in Eq. (6), one may
calculate the matter-corrected neutrino masses $\tilde{m}_i$ in an
analytically exact way. The relevant results have been presented
in Refs. \cite{Barger,Xing00}, from which both
$\widetilde{\Delta}_{ij} \equiv \tilde{m}^2_i - \tilde{m}^2_j$ and
$\widehat{\Delta}_{ij} \equiv m^2_i - \tilde{m}^2_j$ appearing in
Eqs. (19)--(22) and (29)--(34) can straightforwardly be read off.
Explicitly, we have
\begin{eqnarray}
\widetilde{\Delta}_{21} & = & \displaystyle
\frac{2}{3} \sqrt{x^2 - 3y}
\sqrt{3 \left (1 - z^2 \right )} \;\; ,
\nonumber \\
\widetilde{\Delta}_{31} & = & \displaystyle \frac{1}{3} \sqrt{x^2
- 3y} \left [3 z + \sqrt{3 \left (1 - z^2 \right )} \right ] \; ,
\nonumber \\
\widetilde{\Delta}_{32} & = & \displaystyle \frac{1}{3} \sqrt{x^2
- 3y} \left [3 z - \sqrt{3 \left (1 - z^2 \right )} \right ] \; ;
\end{eqnarray}
and
\begin{eqnarray}
\widehat{\Delta}_{11} & = & \displaystyle - \frac{1}{3} x +
\frac{1}{3} \sqrt{x^2 - 3y} \left [ z + \sqrt{3 \left (1 - z^2
\right )} \right ] \; ,
\nonumber \\
\widehat{\Delta}_{22} & = & \displaystyle - \frac{1}{3} x +
\frac{1}{3} \sqrt{x^2 - 3y} \left [ z - \sqrt{3 \left (1 - z^2
\right )} \right ] + \Delta_{21} \; ,
\nonumber \\
\widehat{\Delta}_{33} & = & \displaystyle - \frac{1}{3} x -
\frac{2}{3} z \sqrt{x^2 - 3y} ~ + \Delta_{31} \; ;
\end{eqnarray}
together with
\begin{eqnarray}
\widehat{\Delta}_{21} & = & \Delta_{21} + \widehat{\Delta}_{11} \;
,
\nonumber \\
\widehat{\Delta}_{31} & = & \Delta_{31} + \widehat{\Delta}_{11} \;
,
\nonumber \\
\widehat{\Delta}_{32} & = & \Delta_{32} + \widehat{\Delta}_{22} \;
,
\end{eqnarray}
where $\Delta_{ij} \equiv m^2_i - m^2_j$, and $x$, $y$ and $z$ are given by
\begin{eqnarray}
x & = & \Delta_{21} + \Delta_{31} + A \; ,
\nonumber \\
y & = & \displaystyle
\Delta_{21} \Delta_{31} + A \left [
\Delta_{21} \left ( 1 - |V_{e2}|^2 \right )
+ \Delta_{31} \left ( 1 - |V_{e3}|^2 \right ) \right ] \; ,
\nonumber \\
z & = & \displaystyle
\cos \left [ \frac{1}{3} \arccos \frac{2x^3 -9xy + 27
A \Delta_{21} \Delta_{31} |V_{e1}|^2}
{2 \left (x^2 - 3y \right )^{3/2}} \right ] \;
\end{eqnarray}
with $A \equiv 2 E a$ being the matter parameter.

\newpage

\begin{table}
\caption{Numerical illustration of the terrestrial matter effect on
nine inner angles of six effective unitarity triangles
($\tilde{\triangle}_{e,\mu,\tau}$ and $\tilde{\triangle}_{1,2,3}$
in a long-baseline neutrino oscillation experiment, where $\nu$
stands for neutrinos and $\overline{\nu}$ represents antineutrinos.}
\begin{center}
\begin{tabular}{c|cccc}
Angle ~ & $\tilde{V} = V$ & $E=1$ GeV & $E=2$ GeV & $E=3$ GeV  \\ \hline
$\widetilde{\angle} 1$
& $166.9^\circ$
& $\matrix{ 143.3^\circ ~ (\nu) \cr 141.2^\circ ~ (\overline{\nu})}$
& $\matrix{ 104.9^\circ ~ (\nu) \cr 119.0^\circ ~ (\overline{\nu})}$
& $\matrix{ ~73.4^\circ ~ (\nu) \cr 103.2^\circ ~ (\overline{\nu})}$
\\ \hline
$\widetilde{\angle} 2$
& $3.9^\circ$
& $\matrix{ 35.5^\circ ~ (\nu) \cr ~~0.8^\circ ~ (\overline{\nu})}$
& $\matrix{ 74.5^\circ ~ (\nu) \cr ~~0.4^\circ ~ (\overline{\nu})}$
& $\matrix{ 106.1^\circ ~ (\nu) \cr ~~0.2^\circ ~ (\overline{\nu})}$
\\ \hline
$\widetilde{\angle} 3$
& $9.2^\circ$
& $\matrix{ ~~1.2^\circ ~ (\nu) \cr 38.1^\circ ~ (\overline{\nu})}$
& $\matrix{ ~~0.6^\circ ~ (\nu) \cr 60.7^\circ ~ (\overline{\nu})}$
& $\matrix{ ~~0.5^\circ ~ (\nu) \cr 76.6^\circ ~ (\overline{\nu})}$
\\ \hline
$\widetilde{\angle} 4$
& $6.6^\circ$
& $\matrix{ 18.4^\circ ~ (\nu) \cr 19.4^\circ ~ (\overline{\nu})}$
& $\matrix{ 37.6^\circ ~ (\nu) \cr 30.5^\circ ~ (\overline{\nu})}$
& $\matrix{ 53.3^\circ ~ (\nu) \cr 38.4^\circ ~ (\overline{\nu})}$
\\ \hline
$\widetilde{\angle} 5$
& $88.1^\circ$
& $\matrix{ 72.2^\circ ~ (\nu) \cr 89.6^\circ ~ (\overline{\nu})}$
& $\matrix{ 52.7^\circ ~ (\nu) \cr 89.8^\circ ~ (\overline{\nu})}$
& $\matrix{ 36.9^\circ ~ (\nu) \cr 89.9^\circ ~ (\overline{\nu})}$
\\ \hline
$\widetilde{\angle} 6$
& $85.4^\circ$
& $\matrix{ 89.4^\circ ~ (\nu) \cr 71.0^\circ ~ (\overline{\nu})}$
& $\matrix{ 89.7^\circ ~ (\nu) \cr 59.7^\circ ~ (\overline{\nu})}$
& $\matrix{ 89.8^\circ ~ (\nu) \cr 51.7^\circ ~ (\overline{\nu})}$
\\ \hline
$\widetilde{\angle} 7$
& $6.6^\circ$
& $\matrix{ 18.4^\circ ~ (\nu) \cr 19.4^\circ ~ (\overline{\nu})}$
& $\matrix{ 37.6^\circ ~ (\nu) \cr 30.5^\circ ~ (\overline{\nu})}$
& $\matrix{ 53.3^\circ ~ (\nu) \cr 38.4^\circ ~ (\overline{\nu})}$
\\ \hline
$\widetilde{\angle} 8$
& $88.1^\circ$
& $\matrix{ 72.2^\circ ~ (\nu) \cr 89.6^\circ ~ (\overline{\nu})}$
& $\matrix{ 52.7^\circ ~ (\nu) \cr 89.8^\circ ~ (\overline{\nu})}$
& $\matrix{ 36.9^\circ ~ (\nu) \cr 89.9^\circ ~ (\overline{\nu})}$
\\ \hline
$\widetilde{\angle} 9$
& $85.4^\circ$
& $\matrix{ 89.4^\circ ~ (\nu) \cr 71.0^\circ ~ (\overline{\nu})}$
& $\matrix{ 89.7^\circ ~ (\nu) \cr 59.7^\circ ~ (\overline{\nu})}$
& $\matrix{ 89.8^\circ ~ (\nu) \cr 51.7^\circ ~ (\overline{\nu})}$
\end{tabular}
\end{center}
\end{table}

\newpage

\begin{figure}
\vspace{-1cm}
\epsfig{file=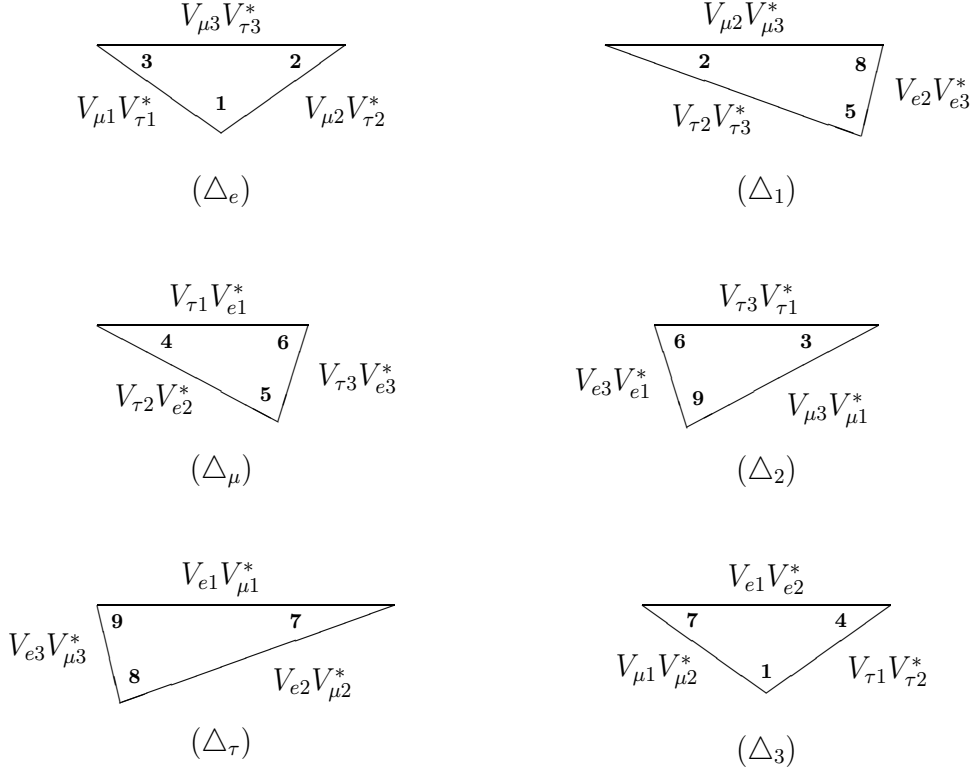,bbllx=2.5cm,bblly=8cm,bburx=18.5cm,bbury=30cm,%
width=15cm,height=22cm,angle=0,clip=}
\vspace{-7.9cm}
\caption{Leptonic unitarity triangles in the complex plane. Each
triangle is named by the index that does not manifest in its three
sides.}
\end{figure}

\newpage

\begin{figure}
\vspace{-1cm}
\epsfig{file=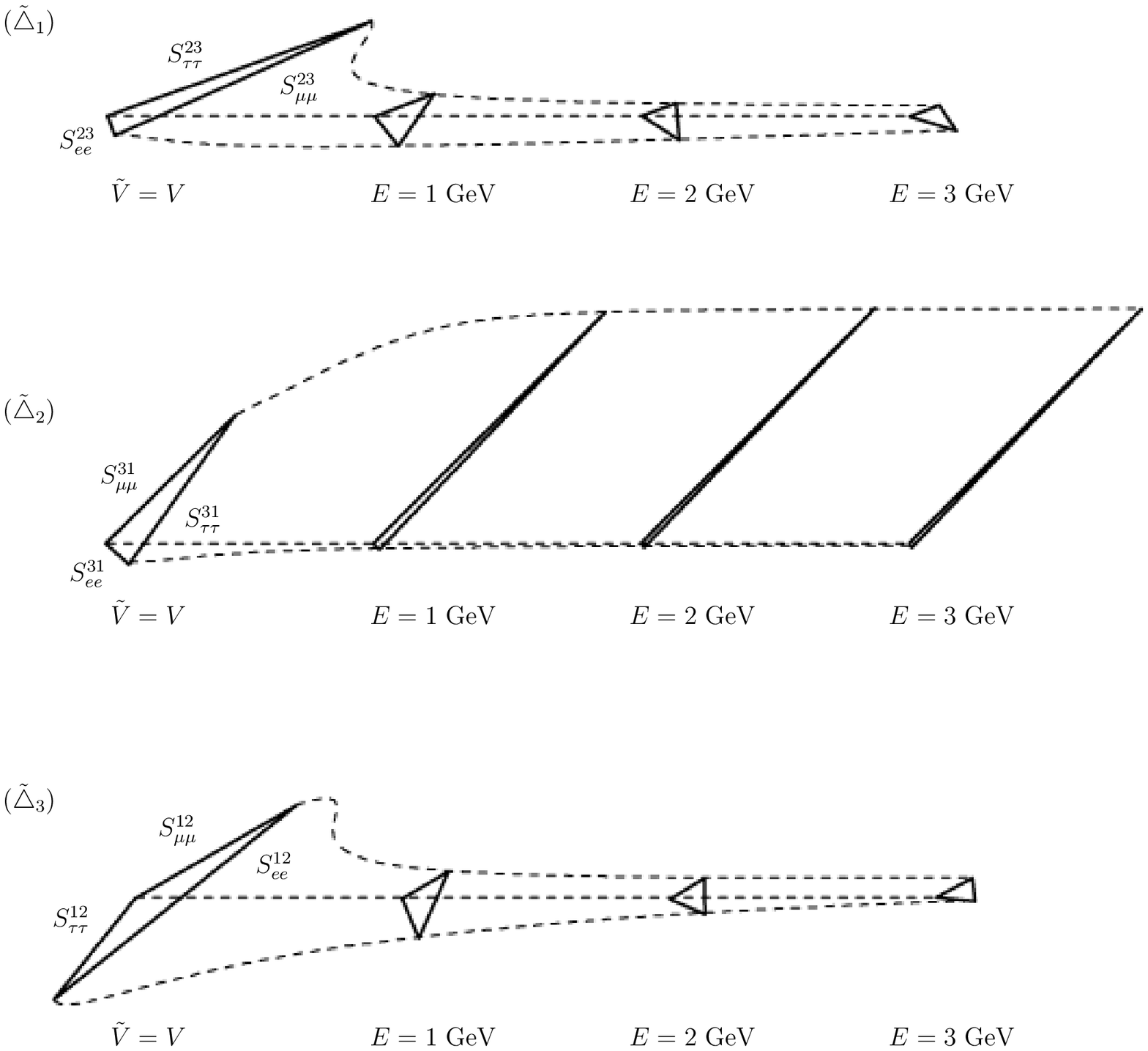,bbllx=0.5cm,bblly=3cm,bburx=20cm,bbury=28cm,%
width=15cm,height=22cm,angle=0,clip=}
\vspace{-4.1cm}
\caption{The shape evolution of three effective
unitarity triangles $(\tilde{\triangle}_1, \tilde{\triangle}_2,
\tilde{\triangle}_3)$ with the beam energy of {\bf neutrinos}
($+A$ and $V$) in a realistic long-baseline oscillation experiment,
where $S^{ij}_{\alpha\alpha} \equiv |\tilde{V}_{\alpha i}
\tilde{V}^*_{\alpha j}|$
(for $\alpha = e, \mu, \tau$ and $i,j = 1,2,3$) has been defined.}
\end{figure}

\newpage

\begin{figure}
\vspace{-1cm}
\epsfig{file=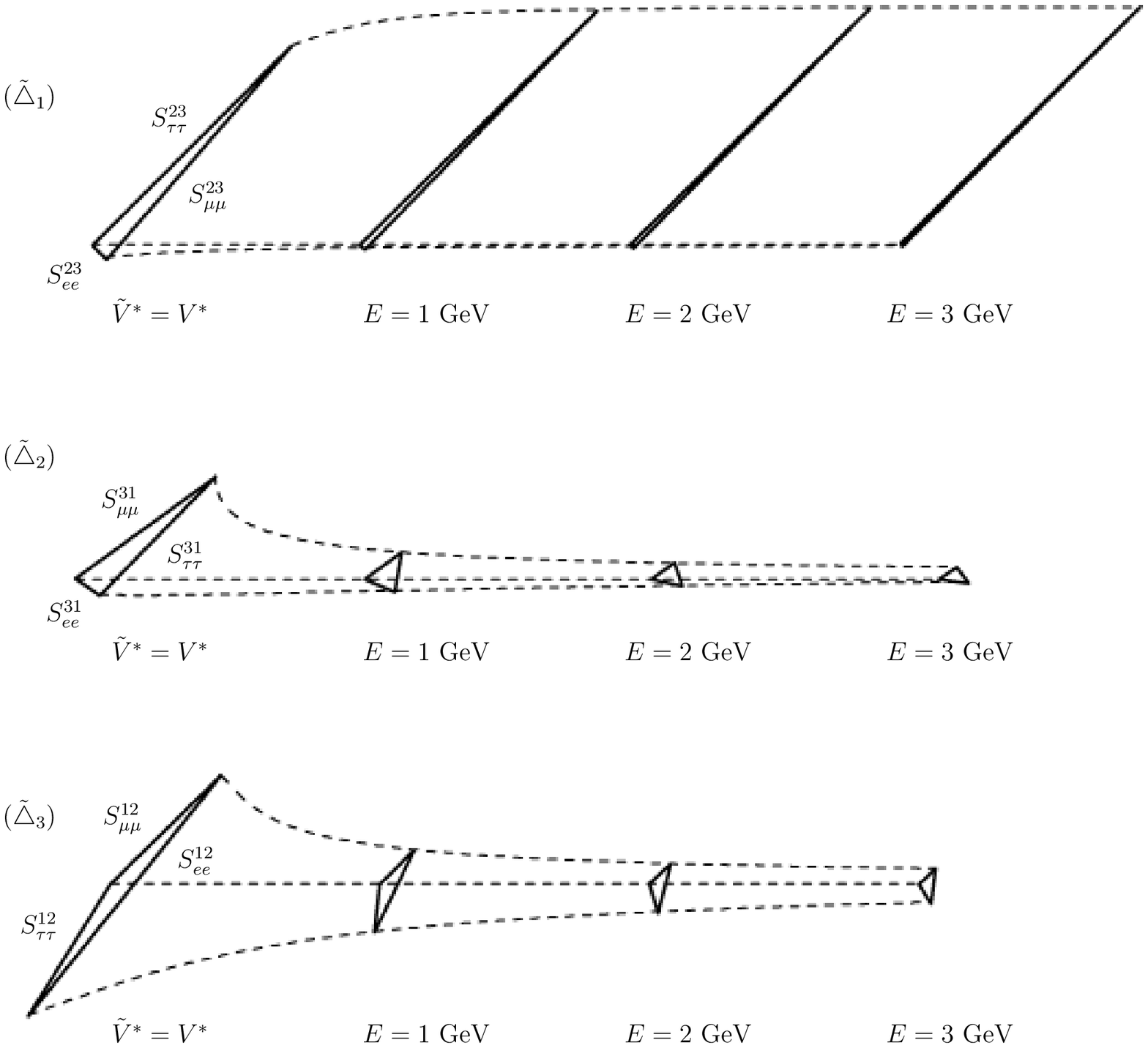,bbllx=0.5cm,bblly=3cm,bburx=20cm,bbury=28cm,%
width=15cm,height=22cm,angle=0,clip=}
\vspace{-4.35cm}
\caption{The shape evolution of three effective
unitarity triangles $(\tilde{\triangle}_1, \tilde{\triangle}_2,
\tilde{\triangle}_3)$ with the beam energy of {\bf antineutrinos}
($-A$ and $V^*$) in a realistic long-baseline oscillation experiment,
where $S^{ij}_{\alpha\alpha} \equiv |\tilde{V}_{\alpha i}
\tilde{V}^*_{\alpha j}|$
(for $\alpha = e, \mu, \tau$ and $i,j = 1,2,3$) has been defined.}
\end{figure}

\newpage

\begin{figure}
\vspace{-1cm}
\epsfig{file=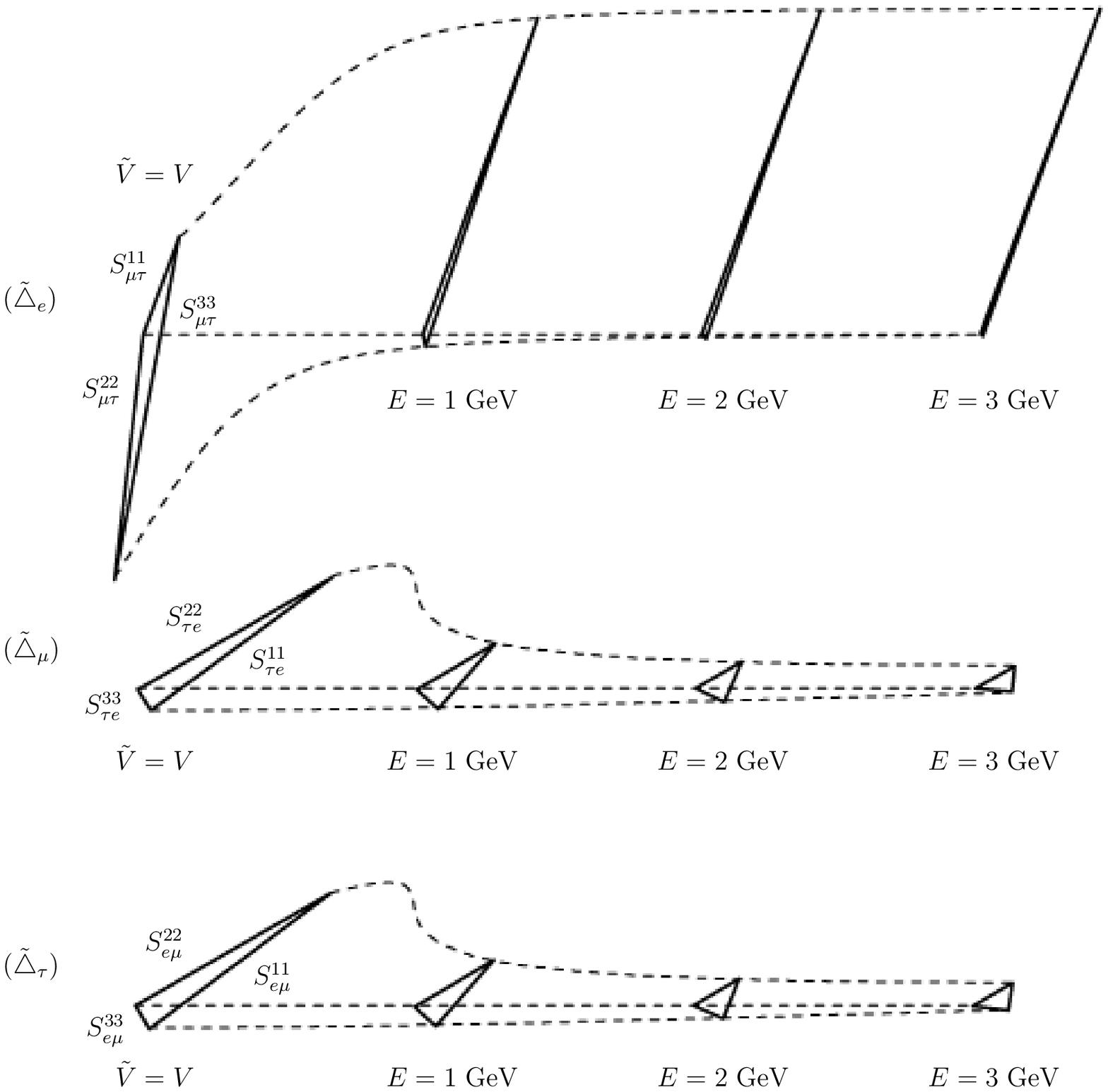,bbllx=0.5cm,bblly=3cm,bburx=19.5cm,bbury=29cm,%
width=15cm,height=22cm,angle=0,clip=}
\vspace{-4.75cm}
\caption{The shape evolution of three effective
unitarity triangles $(\tilde{\triangle}_e, \tilde{\triangle}_\mu,
\tilde{\triangle}_\tau)$ with the beam energy of {\bf neutrinos}
($+A$ and $V$) in a realistic long-baseline oscillation experiment,
where $S^{ii}_{\alpha\beta} \equiv |\tilde{V}_{\alpha i}
\tilde{V}^*_{\beta i}|$
(for $i=1,2,3$ and $\alpha, \beta = e, \mu, \tau$) has been defined.}
\end{figure}

\newpage

\begin{figure}
\vspace{-1cm}
\epsfig{file=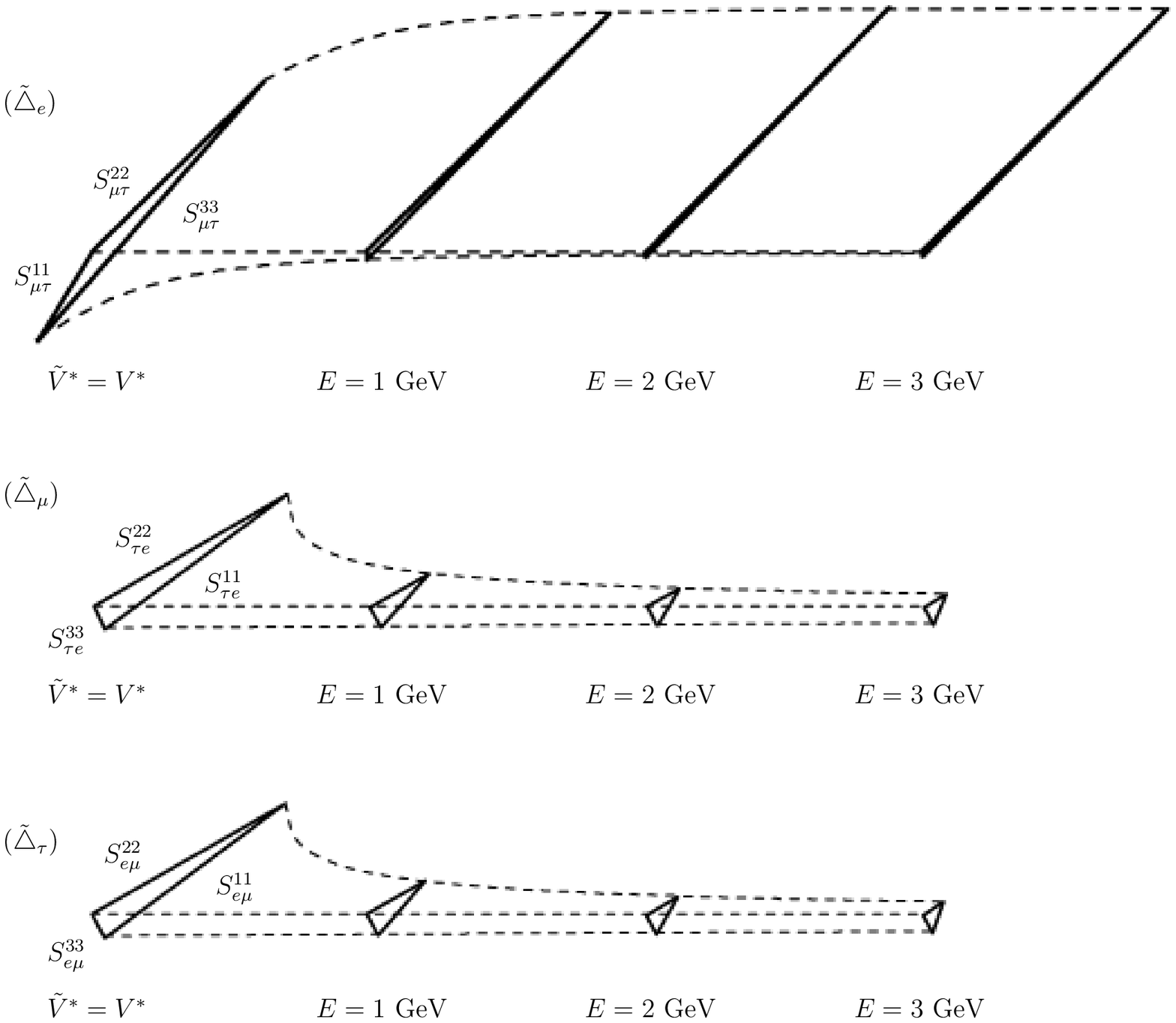,bbllx=0.5cm,bblly=2cm,bburx=19.5cm,bbury=26cm,%
width=15cm,height=22cm,angle=0,clip=}
\vspace{-4.35cm}
\caption{The shape evolution of three effective unitarity
triangles $(\tilde{\triangle}_e, \tilde{\triangle}_\mu,
\tilde{\triangle}_\tau)$ with the beam energy of {\bf
antineutrinos} ($-A$ and $V^*$) in a realistic long-baseline
oscillation experiment, where $S^{ii}_{\alpha\beta} \equiv
|\tilde{V}_{\alpha i} \tilde{V}^*_{\beta i}|$ (for $i=1,2,3$ and
$\alpha, \beta = e, \mu, \tau$) has been defined.}
\end{figure}

\newpage

\begin{figure}
\vspace{-1cm}
\epsfig{file=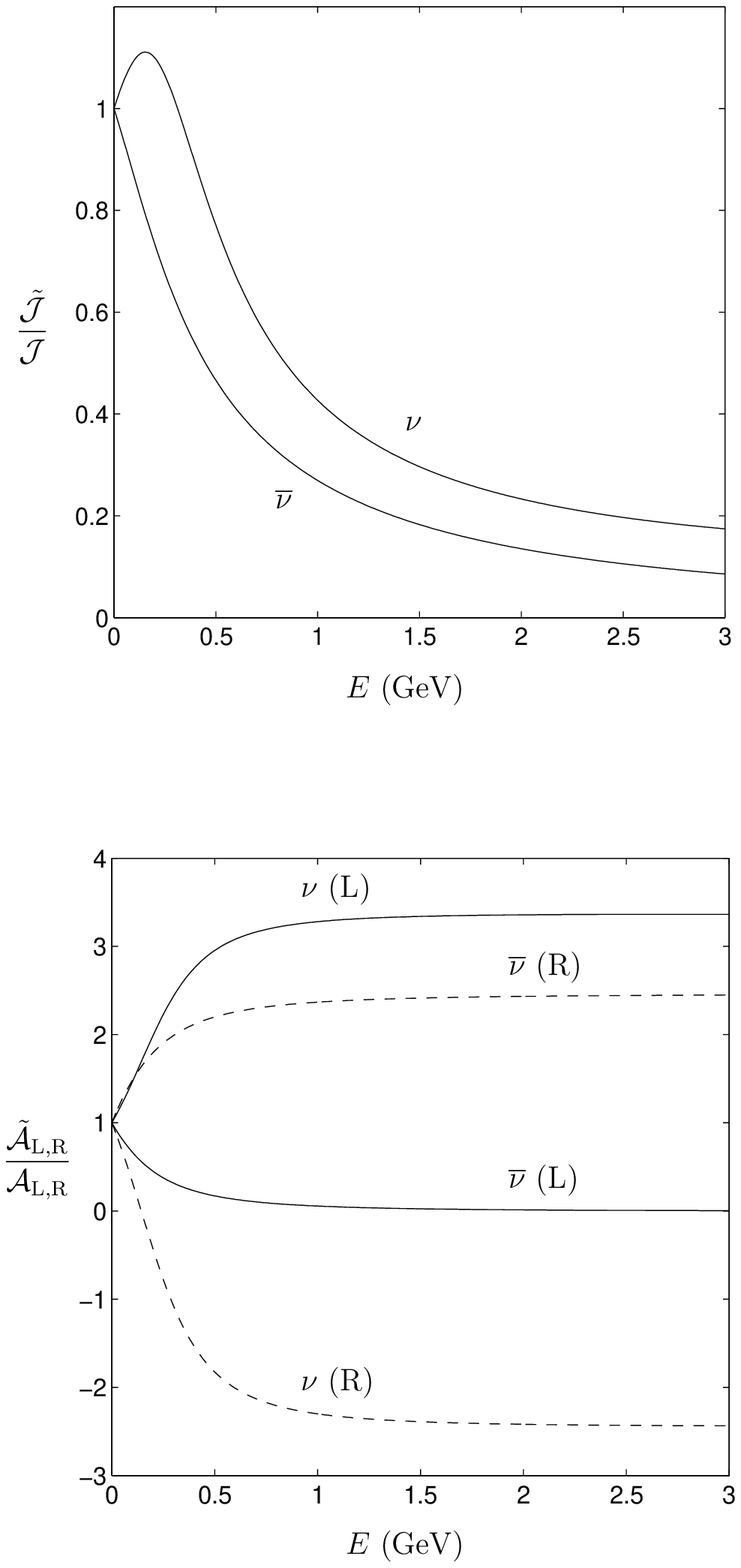,bbllx=1.5cm,bblly=4cm,bburx=19cm,bbury=29cm,%
width=15cm,height=22cm,angle=0,clip=}
\vspace{-1.1cm}
\caption{Terrestrial matter effects on ${\cal J}$, ${\cal A}_{\rm L}$
and ${\cal A}_{\rm R}$ for neutrinos ($\nu$ with $+A$ and $V$) and
antineutrinos ($\overline{\nu}$ with $-A$ and $V^*$) in a
realistic long-baseline oscillation experiment.}
\end{figure}

\end{document}